\documentclass[%
superscriptaddress,
 amsmath,amssymb,
 aps,
 prl,
 reprint,
floatfix,
]{revtex4-2}
\usepackage{graphicx,xcolor}
\usepackage{ulem}
\usepackage{pifont}
\definecolor{darkblue}{RGB}{0,0,150}
\definecolor{nightblue}{RGB}{0,0,100}

\usepackage{mathrsfs,dsfont,mathtools}
\usepackage{dcolumn}
\usepackage{bm}
\usepackage{enumitem}
\usepackage{comment}
\usepackage[
colorlinks,
citecolor=darkblue,
linkcolor=darkblue,
urlcolor=nightblue]{hyperref}

\usepackage[english]{babel}
\usepackage[babel,kerning=true,spacing=true]{microtype}

\usepackage{feynmp-auto}
\usepackage{MnSymbol}

\definecolor{DarkRed}{RGB}{100,0,0}
\definecolor{DarkBlue}{RGB}{000,0,100}

\newcommand{\bscco}{Bi$_2$Sr$_2$CaCu$_2$O$_{8+x}$}

\AtBeginDocument{}

\usepackage{xr}
\makeatletter
\newcommand*{\addFileDependency}[1]{
  \typeout{(#1)}
  \@addtofilelist{#1}
  \IfFileExists{#1}{}{\typeout{No file #1.}}
}
\makeatother
\newcommand*{\myexternaldocument}[1]{%
    \externaldocument{#1}%
    \addFileDependency{#1.tex}%
    \addFileDependency{#1.aux}%
}
\myexternaldocument{letter_SI}

\begin{document}

\title{Moir\'e-induced gapped phases in twisted nodal superconductors
}

\author{Kevin P.~Lucht}
\email{kpl55@rutgers.edu}
\affiliation{%
 Department of Physics and Astronomy, Center for Materials Theory,
Rutgers University, Piscataway, NJ 08854, USA
}%
\author{J.~H.~Pixley}
\affiliation{%
 Department of Physics and Astronomy, Center for Materials Theory,
Rutgers University, Piscataway, NJ 08854, USA
}%
\affiliation{
Center for Computational Quantum Physics, Flatiron Institute, 162 5th Avenue, New York, NY 10010
}%
\author{Pavel A.~Volkov}%
\affiliation{%
Department of Physics, University of Connecticut, Storrs, Connecticut 06269, USA
}%

\begin{abstract}
We demonstrate the emergence of gapped  phases driven by the moir\'{e} superlattice that trivialize the topological states in twisted nodal superconductors. The effect arises from umklapp tunneling between non-adjacent Dirac points in momentum space close to specific twist angles or chemical potentials, determined by the Fermi surface geometry.
We confirm the robustness of the non-topological phase against interactions with self-consistent calculations and show that this gap competes with the previously predicted topological gapped phases, leading to topological phase transitions. These transitions were overlooked in prior literature, signifying the necessity of modifying the phase diagrams of topological phases exhibited in twisted nodal superconductors with and without an interlayer current. We also estimate the relevant twist angles and discuss experimental signatures, focusing on twisted \bscco.
\end{abstract}

\maketitle

%
%

Moir\'{e} materials \cite{balents2020review}, by manipulating the electronic structure, demonstrate the emergence of a 
multitude
of strongly correlated phases, often intertwined with topology. Modeling of these materials generally concentrates on the low energy dispersion centered around high-symmetry points in the Brillouin Zone (BZ). Common platforms include twisted bilayer graphene~\cite{BM2010,Carr2019} and transition metal dichalcogenides (TMDs)~\cite{Angeli2021, Devakul2021,Mao2024, Liu2025}, and heterobilayers such as graphene on h-BN \cite{Jung2014} and MoTe$_2$/WSe$_2$~\cite{Wu2018, Guerci2023, Guerci2024}. In these platforms, moir\'e effects play a prominent role, producing minibands that become flat at specific angles referred to as magic angles~\cite{BM2010, Li2010}.

\begin{figure*}[htbp]
    \centering
\includegraphics[width=1\textwidth]{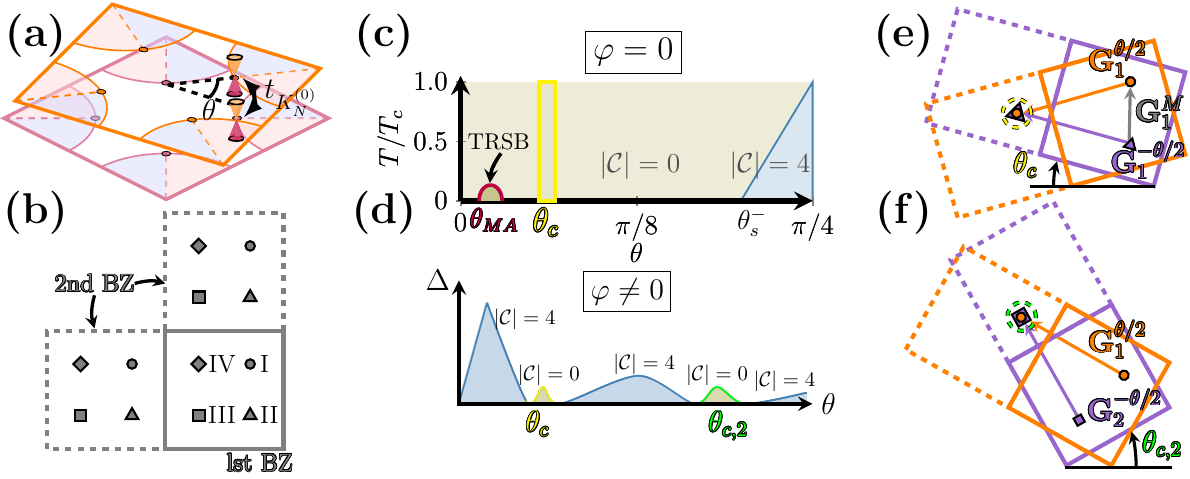}
\caption{{\bf Leading moir\'e umklapp processes in twisted nodal superconductors}: 
(a) schematic of tunneling within a Brillouin zone (BZ) of two layers rotated by an angle $\theta$. (b) shows node's of each valley (labeled as Roman numerals) shown as symbols for each layer. 1st BZ is solid while 2nd BZ is dashed. 
(c) Phase diagram of the superconducting state under twist (up to $\theta = 45^\circ$) and temperature (up to $T_c$) with no interlayer current ($\varphi = 0$). A new non-topological gapped $d$-wave state appears represented by the yellow region. (d) shows energy gap versus matching twist angle range as (c) induced by interlayer current ($\varphi \neq 0$). Topological phase transitions appear in yellow and green regions. Color fill in (c,d) indicates Chern number.
(e,f) show intervalley umklapp tunneling for layers colored in purple and orange. (e) corresponds to yellow phase at $\theta_c$ in (a,b) from valleys $I$ and $II$ related by a moir\'{e} reciprocal lattice vector, $\bm{G}_1^M$. (f) shows green phase at $\theta_{c,2}$  in (b) from valleys $I$ and $III$ caused by differing reciprocal lattice vectors from the two layers.
}
    \label{fig:figure1}
\end{figure*}

More recently, the ideas of twistronics have been applied to twisted nodal superconductor (TNS) interfaces and multilayers~\cite{tns_review}, see Fig.~\ref{fig:figure1} (a). Emergent time-reversal symmetry breaking (TRSB) phases have been predicted both at low~\cite{pavel_article_2023,tummuru2022_2} and large (45$^\circ$ for d-wave SCs)~\cite{Can2021,volkov_jos_2025,Song2022} twist angles (depicted in Fig.~\ref{fig:figure1} (c) ). These phases have been found to be fully gapped and, in most cases, have nontrivial topology~\cite{Can2021,Song2022,pavel_letter_2023}. A topological phase has also been shown to be induced by current at any twist angle~\cite{pavel_letter_2023}.
Remarkably, the importance of minibands and the reconstruction of the BZ by moir\'e superlattice has been much less prominent in the present phenomenology  of TNSs~\cite{tns_review,confalone2025challenges}. There, the Bogoliubov-de Gennes (BdG) quasiparticle Dirac points occur at generic points inside the Brillouin zone (Fig.~\ref{fig:figure1} (a)), and while a twist induces the flattening of the dispersion and can facilitate opening of topological gaps~\cite{Can2021,pavel_letter_2023}, no isolated minibands form~\cite{pavel_article_2023}. Moreover, special attention of this field has been focused on TRSB near $45^\circ$ twist in d-wave superconductors (Fig.~\ref{fig:figure1} (c))~\cite{Yang2018,Can2021,xue_2021,zhang2023josephson,xue_2023_OP,Song2022,Lu2022,Tummuru2022,Haenel2022,Fidrysiak2023,zhou2023,lee2021twisted,Zhao2023,lee2023encapsulating,martini2023twisted,Guo2025,volkov_jos_2025}, where the underlying physics can to a large extent be understood from the point group symmetry properties alone~\cite{Can2021,Volkov2024,tns_review}.  

In this work, we demonstrate that moir\'e superlattice effects, sketched in Fig.~\ref{fig:figure1}(e,f) lead to the emergence of gapped, non-topological phases in twisted nodal superconductors, trivializing the underlying topological phase. They occur in the vicinity of twist angles where Dirac cones, that originate from different valleys in the first BZ, overlap in higher BZs due to umklapp scattering.

Higher BZ overlap can lead to strong moir\'e umklapp tunneling (Fig.~\ref{fig:figure1} (e,f)) and as a result, we find two types of new phases with Chern number $\mathcal{C} =0$ as a function of the twist angle. One emerges in the vicinity of a specific twist angle $\theta_c$ (Fig.~\ref{fig:figure1} (c)) and is expected to persist at all temperatures. In the presence of interlayer current, (Fig.~\ref{fig:figure1} (d)), this $\mathcal{C} =0$ phase competes with the topological ones and we find an additional $\mathcal{C} =0$ phase emerges around another angle $\theta_{c,2}$, leading to a sequence of topological phase transitions as a function of twist angle. We demonstrate that the predicted $\mathcal{C}=0$ phases are distinct from the ones previously predicted in the vicinity of $45^\circ$~\cite{Can2021,Song2022} within a continuum theory and while retaining the full self-consistent order parameter calculations. We finally discuss $\theta_{c},\theta_{c,2}$ estimates and experimental signatures of the predicted phases for twisted \bscco~(BSSCO).

%
%
\textit{Lattice model and moir\'{e} umklapp---}
To provide a qualitative understanding of the effect of moir\'{e} umklapp on nodal superconductors, we first express a twisted bilayer of $d$-wave superconductors with a mean field Hamiltonian, $\hat{H}_{MFT}$~\cite{Can2021,Lucht2025}. This choice of Hamiltonian will be justified in a following section by treating interactions self-consistently to determine the order parameter. $\hat{H}_{MFT}$ is expressed as:
\begin{equation}
\begin{gathered}
\hat{H}_{MFT}  =  -t \sum_{\langle i,j \rangle, l, \sigma} ( c_{il\sigma}^\dagger c_{jl\sigma} +  \mathrm{h.c.} ) -\mu  \sum_{i,l,\sigma} n_{il\sigma}
\\
+
\sum_{\langle i,j \rangle, l} ( \Delta_{ij}e^{i\varphi_l} c_{il\uparrow}^\dagger c^\dagger_{jl\downarrow} + \mathrm{h.c.})
-\sum_{ i,j ,\sigma} g(r_{ij}) ( c_{i1\sigma}^\dagger c_{j,2,\sigma} +\mathrm{h.c.}),
\label{eqn:Hlat}
\end{gathered}
\end{equation}
where indices $i,j$ denote individual layer sites (in cartesian coordinates aligned with each layer), while $l=1,2$ - layers. The interlayer tunneling is given by
\begin{equation}
    g(r_{ij}) = g_0 e^{-\frac{{\sqrt{r^2_{ij} - d^2} - d}}{\rho}},
    \label{eq:num_tun_strength}
\end{equation}
where $g_0$ is the bare tunneling strength, $r_{ij}$ is the in-plane distance between sites $i$ and $j$ in two twisted layers, $d$ is the interlayer spacing, and $\rho$ sets the tunneling range. To perform a twist, we chose commensurate twist angles $\theta_{nm}=2\tan^{-1}(n/m)$ where $n,m \in \mathbb{Z}$~\cite{Can2021}. Details of the Hamiltonian are provided in SM Appendix I~\cite{SM}.

We begin by assuming a d-wave order parameter, i.e. $\Delta_{ij}$ to be equal to $+(-)\Delta_0/2$ for bonds along $x(y)$. In momentum space, this leads to $\Delta_{\bm{k}} = \gamma_{\bm{k}} \Delta_0/2$ where $\gamma_{\bm{k}}= \cos k_x - \cos k_y$ in local coordinates of each layer. The four Dirac points (Which we will refer to below as valleys, and labeled directly in Fig.~\ref{fig:figure1} (b) with different shaped symbols) of the resulting quasiparticle dispersion, Fig.~\ref{fig:figure1} (a), are thus displaced in two layers. 
Continuing Fig.~\ref{fig:figure1} (a) beyond the first BZ, however, one finds that different Dirac points from two layers can still overlap in higher BZs (Fig.~\ref{fig:figure1} (b,e,f)). Specifically, for a fixed twist angle there exists a chemical potential value where they overlap already in the second BZ.


The Dirac points overlapping in higher BZs develop an additional resonant coupling, mediated by the effect of the moir\'e superlattice, i.e. {\it moir\'e umklapp} coupling. Indeed, in momentum space, the tunneling takes the form~\cite{BM2010,pavel_article_2023}:
\begin{equation}
    t_{\bm{K},\tilde{\bm{K}}} = \sum_{\bm{G},\tilde{\bm{G}}} \frac{g_{\tilde{\bm{K}}+\tilde{\bm{G}}} }{\Omega} \delta_{\bm{K} + \bm{G},\tilde{\bm{K}}+\tilde{\bm{G}}'}
    \label{eqn:tunneling_form_factor}
\end{equation}
where $g_{\tilde{\bm{K}}+\tilde{\bm{G}}}$ is the Fourier transform of $g(r)$ and $\bm{G}$ and $\tilde{\bm{G}}$ are the reciprocal lattice vectors of the original and twisted BZ, respectively. As a result, quasiparticles at momenta that overlap in higher BZs ($\bm{G},\tilde{\bm{G}} \neq 0$) still couple, albeit with a reduced amplitude. Such coupling has been studied in twisted graphene, where it leads to enhanced interlayer conductivity at special twist angles~\cite{BM2010,mele2010,Chari2016,Koren2016,Inbar2023}, and a hierarcy of minibands can be generated~\cite{fu2020magic}. Similar effects can also occur in topological surface states in the presence of a moire potential~\cite{PhysRevB.103.155157,zhang2025fabrication} and graphene with patterned dielectric superlattices~\cite{forsythe2018band}. Furthermore, quasicrystal structures in twisted bilayer graphene~\cite{Yu2019QCG,Deng2020,Yao2018,Ahn2018}, twisted trilayer graphene~\cite{Uri2023}, twisted WSe$_2$ ~\cite{Li2024}, and MoS$_2$~\cite{Tsang2024} can be modeled by  dominant umklapp scattering processes~\cite{Moon2019}. We will show below that the consequences of this coupling in TNS are more drastic. 


\begin{figure}[h]
    \centering
\includegraphics[width=.48\textwidth]{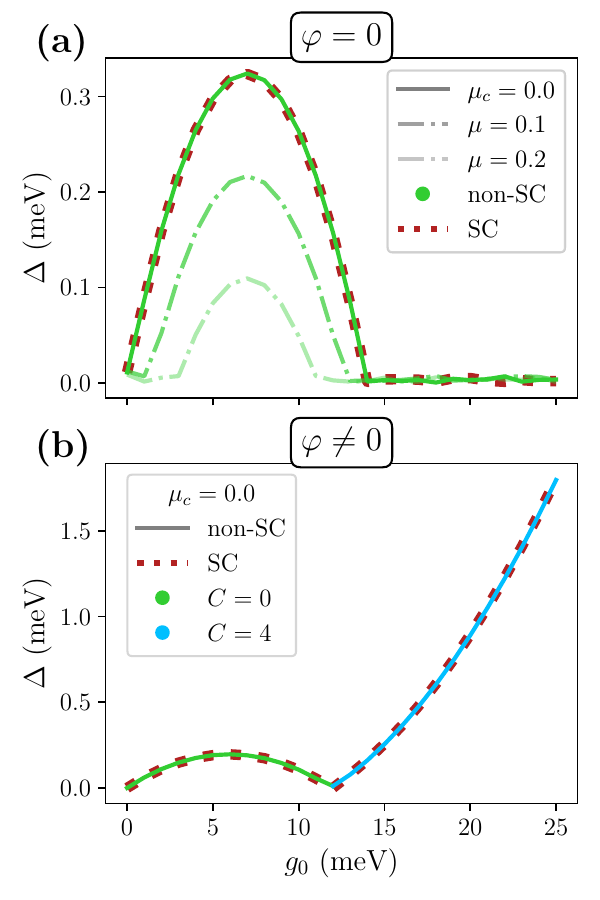}
\caption{{\bf Gap opening due to moir\'e umklapp near $\mu=\mu_{c}$}: Displayed are the energy gap, labeled by $\Delta$, for varying tunneling strength, $g_0$, and for chemical potential $\mu$ (in units of meV) around $\mu_c$ (see Fig.~\ref{fig:figure1} (e)). (a) and (b) 
are for phase differences of $\varphi=0$ and $\varphi=\pi/2$, respectively.  For $\varphi=0$, a non-topological gap ($|\mathcal{C}|=0$) opens in the spectrum for $\mu=\mu_c$ and decreases away from $\mu_c$. Once a phase difference is present, $\mu_c$ maintains a non-topological gap for low $g_0$, but a topological gap ($|\mathcal{C}|=4$) opens for large values of $g_0$. Self consistent results (red dashed line) are taken at $\mu_c$ while non-self-consistent results (green and blue solid and dashed lines) use a $d$-wave order parameter with  a magnitude of $40$ meV. Behavior of the gaps for $\varphi=\pi/2$ under varying $\mu$ are shown in the End Matter for $\theta_{1,2}$.
} 
    \label{fig:figure2_n1m3}
\end{figure}

\begin{figure}[h]
    \centering
\includegraphics[width=.48\textwidth]{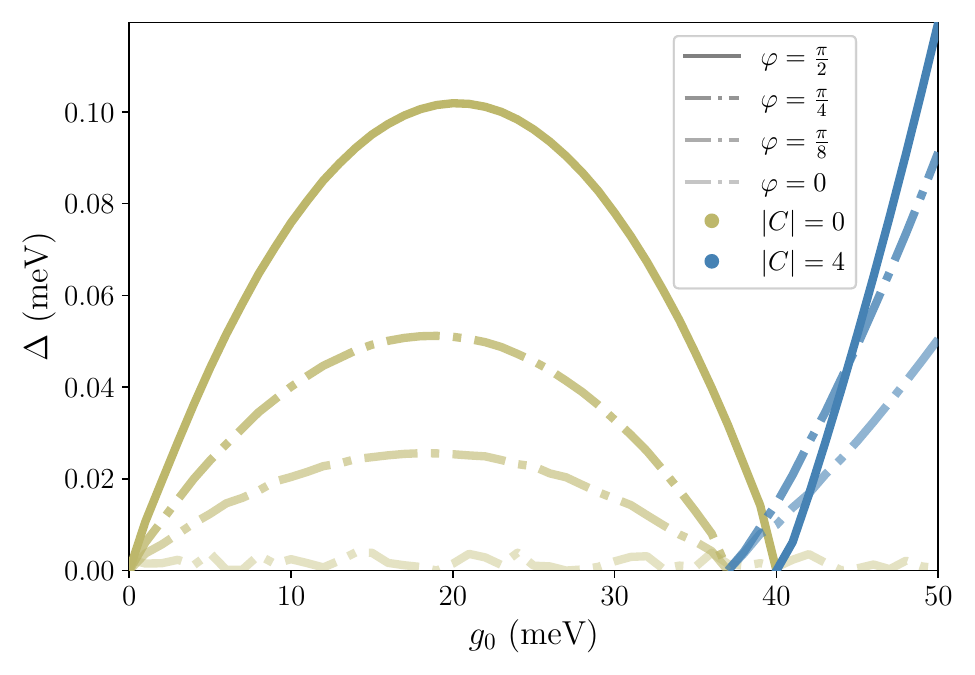}
\caption{{\bf Moir\'e induced gap opening near $\mu=\mu_{c,2}$ with a finite phase difference}: Displayed are the energy gap, $\Delta$, for varying $g_0$ at $\mu_{c,2}$. With no phase difference, the system remains nodal for all $g_0$. Once a phase difference is applied, the gap gradually opens, maximizing at $\varphi = \frac{\pi}{2}$. Results reflect non-self-consistently determined order parameter of a $d$-wave order parameter with a magnitude of $40$ meV. 
} 
\label{fig:figure3_n1m3_muc2}
\end{figure}

\textit{Non-topological gap opening---} We now consider the quasiparticle spectrum of Eq. \eqref{eqn:Hlat} in the presence of the moir\'e umklapp. We will focus on overlap of Dirac points of two layers occurring in the second BZ (see Fig.~\ref{fig:figure1} (e,f) ), as the tunneling strength, Eq.~\eqref{eqn:tunneling_form_factor}, rapidly decays with momentum~\cite{BM2010,pavel_article_2023}. Following Eq.~\ref{eqn:tunneling_form_factor}, the overlap condition is $\bm{K}_N^{\theta/2,(1),I} + \bm{G}^{\theta/2}_i = \bm{K}_N^{-\theta/2,(1),J} + \bm{G}^{-\theta/2}_j$ where $i,j = 1,2$ correspond to reciprocal lattice vectors of two layers and $\bm{K}_N$ is the momentum of a valley's nodal point for a given layer with the superscripts corresponding to twist angle of the layer, BZ, and valley, respectively. The superscript $I,J=0,1,2,3$ enumerates the $4$ different BdG Dirac points of each layer (``valleys"), denoted I, II, III, IV, respectively, with unique symbols in Fig.~\ref{fig:figure1} (b). For $i=j$ (and $J=I+1$ mod $4$), this condition implies that the connected Dirac points are separated by a moir\'{e} reciprocal lattice vector $\bm{G}_i^m =\bm{G}^{\theta/2}_i - \bm{G}^{-\theta/2}_j$.

Rather than fixing $\mu$ and tuning the angle, we equivalently fix the angle and tune the chemical potential. The angles referenced in Fig.~\ref{fig:figure1} (e,f) will therefore have a corresponding chemical potential $ \mu_c \equiv \theta_c$ and $ \mu_{c,2} \equiv \theta_{c,2}$ that produce these conditions. We identified two commensurate twist angles, $\theta_{1,2}$ and $\theta_{1,3}$, where intervalley tunneling between adjacent valleys $I$ and $J=(I\pm1) \mod 4$ occurs for  $\mu_c \approx 0.0$ meV. For $\theta_{1,3}$ as presented in Fig.~\ref{fig:figure1} (e) (valleys labeled by Roman numerals in Fig.~\ref{fig:figure1} (b)), we can also find a separate value $\mu_{c,2} = 2t$, where opposing valleys  $I$ and $J=(I\pm2) \mod 4$ overlap. 
For calculations, we use $t=-126$ meV, $\rho = 2.11$ \AA, $\Delta_0 = 40$ meV  and  $d =  d_0/5$ for $\mu_c$ and $d =3d_0/4 $ for $\mu_{c,2}$, where $d_0 = 12$ \AA~to match previous literature~\cite{Markiewicz2005,Can2021,Lucht2025}. The reduced interlayer distance is taken to amplify the moir\'e umklapp processes, reducing the suppression of $g_{\tilde{\bm{K}}+\tilde{\bm{G}}}$ in Eq. \eqref{eqn:tunneling_form_factor} with $\tilde{\bm{G}}$ (see also Fig.~\ref{fig:figure1} (a)).

Fig.~\ref{fig:figure2_n1m3} (a) and Fig.~\ref{fig:figure3_n1m3_muc2} and plot the gap of the BdG dispersion as $g_0$ is increased for twist angle $\theta_{1,3}$ at $\mu_c$ and $\mu_{c,2}$, respectively (plots for $\theta_{1,2}$ at $\mu_c$ are provided in the End Matter). Surprisingly, $\mu_c$ and $\mu_{c,2} $ can produce qualitatively different band structures. For $\mu_c$, a non-topological gap opens in the spectrum while for $\mu_{c,2}$, the system remains nodal (see Fig.~\ref{fig:figure1} (c) ).

The opening of the gap in the Dirac spectrum at $\mu_c$ raises the question of the topological character of this state. We find that the resulting gapped state is characterized by a Chern number $|\mathcal{C}|=0$ already at the level of individual valleys. To investigate this further, we introduce a phase difference between the layers' order parameters, related to interlayer current via the Josephson relation~\cite{kleiner1994,yurgens2000intrinsic,pavel_article_2023}, which is expected to open a topological gap forming a topological superconductor with $|\mathcal{C}|=4$ \cite{pavel_article_2023,pavel_letter_2023,Lucht2025}. As shown in Fig.~\ref{fig:figure2_n1m3} (b), once the interlayer current is applied, the $|\mathcal{C}|=0$ gap persists at low $g_0$ for $\mu_c$, then closes and reopens at larger $g_0$ now with $|\mathcal{C}|=4$. This confirms that the $|\mathcal{C}|=0$ state produced by the moir\'e umklapp competes with the topological current-induced phase. As shown in Fig.~\ref{fig:figure3_n1m3_muc2}, unexpectedly, a $|\mathcal{C}|=0$ gap opens at low $g_0$ for $\mu_{c,2}$ as well for $\varphi \neq 0$. Thus, the moir\'e umklapp leads to a change of the topological character of the current-induced state (see Fig.~\ref{fig:figure1} (d) for the corresponding phase diagram).

\textit{Self-consistent results---} The results above assumed the order parameters of the two layers to be $d$-wave with a fixed magnitude. However, previous work has pointed out the possibility of phase transitions induced by twist in self-consistent calculations \cite{pavel_article_2023,senechal_2022,Can2021,senechal_2024,senechal_2024_2}, and specifically the emergence of spontaneous TRSB for twists close to $45^\circ$ \cite{Can2021,Zhao2023,volkov_jos_2025,senechal_2022,senechal_2024,senechal_2024_2} as shown in Fig.~\ref{fig:figure1} (c). To study the impact of these interaction effects on the moir\'e umklapp-induced phases, we determine the order parameter in Eq.~\eqref{eqn:Hlat} self-consistently for a model of an attractive nearest-neighbor density-density interaction $H_{int}  = -\sum_{\langle ij\rangle,\sigma\sigma'}  V n_{i\sigma}n_{j\sigma'}$ \cite{Can2021} (see details in SM Appendix I~\cite{SM} and End Matter). For individual layers, the available channels for superconductivity of $H_{int}$ are extended $s$-wave, $d$-wave, or their combination $d+is$. As $g_0$ is turned on, we expect, for at least small $g_0$, that the phase diagram remains relatively consistent. Indeed, we find that a TRSB order parameter remains near the region where $d$-wave crosses over to extended $s$-wave, but now also permits a $d+id$ state which corresponds to the formation of a topological superconductor~\cite{Can2021}. A full illustration of the stable order parameter for varying chemical potential for a single layer and twist angles $\theta_{1,2}$ and $\theta_{1,3}$ is provided in the End Matter.

Near $\mu_c$, however, the order parameter remains $d$-wave. To ensure the stability of the $d$-wave order parameter at $\mu_c$, we perform self-consistency for varying $g_0$ at $\mu_c$ for both $\theta_{1,2}$ and $\theta_{1,3}$.  In either case, the $d$-wave solution remains stable up to $g_0 = 40$ meV, establishing that the non-topological gap formed from moir\'{e} effects is indeed present. Fig.~\ref{fig:figure2_n1m3} displays the resulting energy gap for varying $g_0$ at $\mu_{c}$ for the self-consistently determined order parameter. Thus, the moir\'e umklapp mechanism for the formation of the $|\mathcal{C}|=0$ state is distinct from the previous suggestions~\cite{Can2021} based on the formation of an $d+is$ state. For $\mu_{c,2}$, however, the stable order parameter is extended $s$-wave and therefore moir\'{e} effects will not be present for twist $\theta_{1,3}$. 



%
%

\textit{Continuum model---}
To understand the results of the previous section, we develop a simplified continuum model of the two Dirac nodes in separate valleys and layers coupled via the moir\'e umklapp. We start with an effective Hamiltonian for a single Dirac node in valley $i=1,2,3,4$, including the effects of interlayer coupling within the first BZ perturbatively \cite{pavel_article_2023}:
\begin{equation}
H_i(\bm{k})  =     {\bf v}^{(i)}_F \cdot {\bf k}\tau_3 + ({\bf v}^{(i)}_\Delta \cdot {\bf k} +m_i) \tau_1 \cos \frac{\varphi}{2}+  (m_i-{\bf v}^{(i)}_\Delta \cdot {\bf k} )\tau_2 \sin \frac{\varphi}{2},
\label{eq:intra_Ham}
\end{equation}
where $\tau_i$ are Pauli matrices in Gor'kov-Nambu space, ${\bf v}^{(i)}_F$ is the Fermi velocity, ${\bf v}^{(i)}_\Delta$ is the order parameter's velocity, and  $m =\frac{t_{K_N^{(1)}}^2}{v_\Delta Q_N}$, where $t_{K_N^{(1)}} \ll v_\Delta K_N \theta$ and $Q_N$ is the distance between adjacent Dirac points of two layers in momentum space, equal to $K_N \theta$ at $\theta\ll1$. One can show that for finite $\varphi$ this Hamiltonian reproduces the correct value of the topological gap \cite{pavel_letter_2023}.

\begin{figure}[htbp]
    \centering
\includegraphics[width=.49\textwidth]{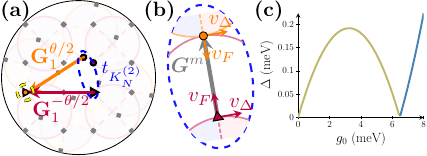}
\caption{{\bf Description of the moir\'e umpklapp process using a continuum model.} (a) Depicts the intervalley tunneling (circled in dashed blue) processes of Fig.~\ref{fig:figure1} (e) with corresponding non-topological energy gap in (c).  (b) shows the intervalley tunneling processes with the orientation of the velocities of each node. (c) illustrates the energy gap versus tunneling strength for $\varphi>0$. Green represents the $|\mathcal{C}|=0$ phase for low $g_0$ while blue corresponds to the topological phase $|\mathcal{C}|=1$ for the two nodes.}
    \label{fig:cont_model}
\end{figure}

We now include intervalley tunneling $t_{K_N^{(2)}}$ for two simplified cases. For coupling between adjacent nodes ($J=[I\pm 1]\mod 4$, see Fig.~\ref{fig:figure1} (e) ), one notices that ${\bf v}^{(I)}_F \approx - {\bf v}^{(J)}_F$ and ${\bf v}^{(I)}_\Delta \approx {\bf v}^{(J)}_\Delta$, resulting in $m_I=m_J$. This scenario is illustrated in Fig.~\ref{fig:cont_model} (a) and (b). The Hamiltonian including the coupling between adjacent valleys, $H_{ANT}(\bm{k})$, then takes the form:
\begin{equation}
\begin{gathered}
H_{ANT}(\bm{k}) = 
v_F k_y \tau_3 \eta_3 +(v_\Delta k_x+m) \cos (\varphi/2) \tau_1 
\\
+ (m-v_\Delta k_x) \sin(\varphi/2) \tau_2 \eta_3
+t_{K_N^{(2)}} \tau_3 \eta_1,
\end{gathered}
\end{equation}
where $\eta_i$ acts in valley space and we choose $k_y \parallel \bm{K}^{(1)}_N$ and $k_x \perp \bm{K}^{(1)}_N$. One finds that for $\varphi=0$ there is always a $2 t_{K_N^{(2)}} $ gap in the spectrum. Note that even in the absence of pairing ($\tau_1$ term) there is a full gap; this indicates that the moir\'e umklapp simply shifts the Fermi surface away from the nodal line. Thus, the remaining Fermi surface does not cross the nodal line, resulting in a ``nodeless d-wave" pairing, somewhat similar to the nodeless $s_\pm$ state in iron-based superconductors \cite{hirschfeld2011gap}. For finite $\varphi$, a topological transition where gap closes and reopens around one Dirac node occurs at  $2m\sin(\varphi/2) = t_{K_N^{(2)}} $, resulting in a change in the Chern number, $\Delta \mathcal{C}$, of $|\Delta \mathcal{C}| =1$.
For the tunneling values of $\theta_{1,2}$ (see SM Appendix VIII for details and values~\cite{SM}), we plot the gap and corresponding topological phase in Fig.~\ref{fig:cont_model} (c) which is fully consistent with Fig.~\ref{fig:figure2_n1m3}. Note that since $m\propto g_0^2$, at low $g_0$ one always has $2m\sin(\varphi/2) < t_{K_N^{(2)}}$. Thus, despite the suppressed amplitude for the moir\'{e} umklapp tunneling, for twist angles realizing direct overlap it is the dominant tunneling process in the weak tunneling limit.

For coupling between opposite nodes $J=[I\pm 2]\mod 4$ (see Fig.~\ref{fig:figure1} (f)), ${\bf v}^{(I)}_F \approx - {\bf v}^{(J)}_F$ and ${\bf v}^{(I)}_\Delta \approx -{\bf v}^{(J)}_\Delta$ and $m_I=-m_J$ (note that $\varphi$ also change sign between two valleys since they come from different layers), resulting in the Hamiltonian, $H_{ONT}(\bm{k})$, expressed as:
\begin{equation}
\begin{gathered}
H_{ONT}(\bm{k}) = 
v_F k_y \tau_3 \eta_3 +(v_\Delta k_x+m) \cos (\varphi/2) \tau_1 \eta_3
\\
+ (m-v_\Delta k_x) \sin(\varphi/2) \tau_2
+t_{K_N^{(2)}} \tau_3 \eta_1.
\end{gathered}
\end{equation}
For $\varphi=0$ there is no gap for any $m,t_{K_N^{(2)}} $. The Dirac points are at $v_\Delta k_x = \pm t_{K_N^{(2)}}-m$. For nonzero $\varphi$, on the contrary, there is always a gap, but it closes and reopens at $2 m \cos(\varphi/2) = t_{K_N^{(2)}}$. Thus, a non-topological gap at low $m$ only opens when TRS is broken in this case, again in agreement with Fig.~\ref{fig:figure2_n1m3}.

The results above for the idealized cases of (anti-)parallel ${\bf v}_F$ and  ${\bf v}_\Delta$ for two Dirac valleys should withstand a finite misorientation due to gapped character of the resulting phases, explaining their observation in the lattice model.

\textit{Discussion---}
Our results demonstrate the emergence of a gapped non-topological phase at specific twist angles in twisted nodal superconductors, which would have been topological if umklapp processes were ignored. The signatures of the gap can be observed in tunneling spectroscopy experiments. A more interesting implication may be that the opening of such a gap will suppress the nodal quasiparticles that lead to a reduction of the superfluid stiffness \cite{leewen}, increasing the resulting critical Berezeinskii-Kosterlitz-Thouless temperature in a 2D superconductor, such as TNSs. Specifically, in underdoped cuprates, where phase fluctuations have been suggested to be extremely large \cite{emery1995importance}, this effect may lead to a considerable enhancement of $T_c$. 

To realize these higher BZ tunneling processes, BSCCO is a likely candidate as it can be exfoliated to the few-layer limit~\cite{Yu2019,Zhao2019}. Here, at a given chemical potential, we determine the critical angle, $\theta_c$, where overlapping nodes in the extended BZ form. We start with a one-band approximation of the intralayer hopping of the normal state (see End Matter for details), and fit remaining parameters and chemical potential with ARPES data~\cite{Markiewicz2005}. Using the known $d_{x^2-y^2}$-wave order parameter of BSCCO~\cite{Tsuei2000}, nodes form along the diagonals of the BZ. At optimal doping, we take $\mu = 130$ meV~\cite{Markiewicz2005} and determine that $\theta_c = 41.1^\circ$. We can also approximate the tunneling strength given the lattice constants and ARPES measurements of $v_\Delta = 43.9$ meV taken near the node~\cite{He2018}. Taking the bare interlayer tunneling strength as $g_0 = 1$ meV~\cite{Markiewicz2005,pavel_letter_2023} and estimating $\delta_0 = v_\Delta | K^{\theta_c/2} -  K^{-\theta_c/2}|$, we find $t_{K_N^{(2)}}/m
\approx 0.93$. For the given Fermi surface, $\theta_{c,2}=30.5^\circ$ is strongly suppressed; details of $\theta_{c,2}$ are provided in the End Matter. Fitting additional ARPES data in the overdoped region also estimates a ratio $\mathcal{O}(1)$ near optimal doping which we provide in the SM Appendix X~\cite{SM}. 
We also note that the tunneling strength is intrinsically angle dependent~\cite{Song2022}, and including an angular form factor further ensures $t_{K_N^{(2)}} > t_{K_N^{(1)}}^2/\delta_0$.

In conclusion, we have demonstrated that a moir\'{e} effect can be found near critical twist angles or chemical potential values where moir\'e umklapp scattering occurs (Fig.~\ref{fig:figure1} (e,f)).  
This effect forms a non-topological gapped states that we verify its stability through self-consistency calculations of the order parameter.
From continuum analytical model, the moir\'e umklapp hybridizes different parts of the Fermi surface, such they no longer cross the nodal line, leading to nodeless $d$-wave behavior.  This moir\'{e}-induced gapped state is distinct from gaps formed from time-reversal symmetry breaking and competes with current-induced topological phases. Thus, previous phase diagrams of TNSs with and without interlayer current require modification.
This moir\'{e} effect has implications for raising $T_c$ and is accessible in the experimental platform of twisted \bscco. 

\acknowledgements{Acknowledgments: This work was partially supported by NSF Grant No.~DMR-2515945 (K.~L.~ and J.~H.~P.). This work was  performed  in part at the Aspen Center for Physics, which is supported by National Science Foundation grant PHY-2210452 (J.~H.~P., P.~A.~V.) and as well as at the Kavli Institute of Theoretical Physics that is supported in part by the National Science Foundation under Grants No.~NSF PHY-1748958 and PHY-2309135 (K.~L., P.~A.~V. and J.H.P.). }

\section{End Matter}

\begin{figure}[h]
    \centering
\includegraphics[width=.5\textwidth]{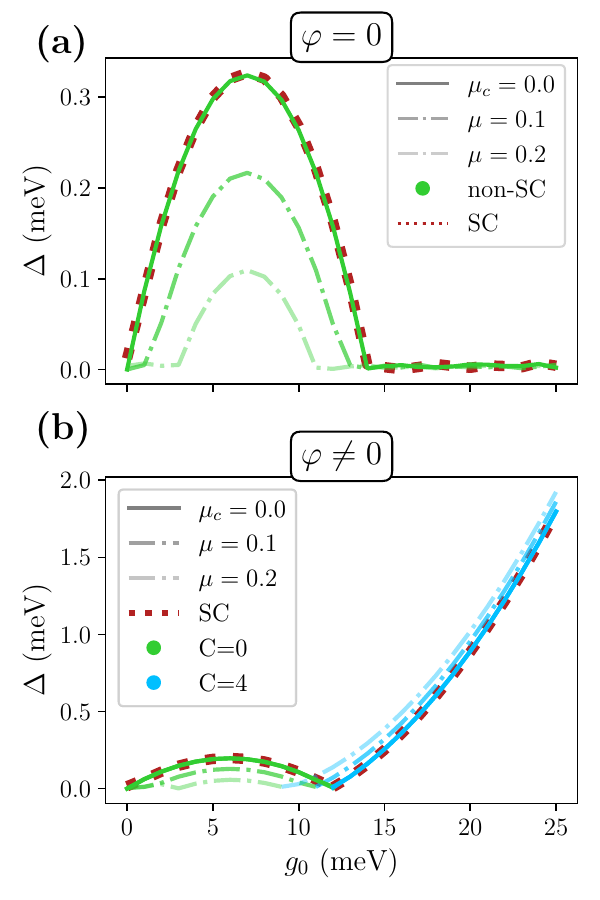}
    \caption{{\bf Energy gap for $\theta_{1,2}$}. We corroborate the gapped spectrum of the $\theta_{1,3}$ structure with $\theta_{1,2}$. Both have a critical chemical potential of $\mu_c = 0.0$ meV. For $\varphi =0$, (a) displays self-consistent and non-self-consistent results at $\mu_c$ and away from $\mu_c$ for non-self-consistent results. (b) shows the gap in the presence of $\varphi \neq 0$ at and near $\mu_c$ without self-consistency and at $\mu_c$ with self-consistency. For low $g_0$, both cases display a non-topological gap. Away from $\mu_c$, the non-topological gap gradually vanishes, leaving only the topological gap at higher $g_0$. Units of $\mu$ given in meV.}
    \label{fig:SI_n1m2_nonSC_gap}
\end{figure}

\textit{Non-self consistent results for $\theta_{1,2}$}---
We note that for the $\theta_{1,2}$, the twist angle exceeds $45^\circ$ which causes a relative $\pi$-phase shift between the layers' order parameters in order to stabilize the free energy. For $\theta_{1,2}$, $\mu_c = 0.0$ meV just as in the case for $\theta_{1,3}$. The gap with and without a phase difference is displayed in Fig.~\ref{fig:SI_n1m2_nonSC_gap} which mimics the results of $\theta_{1,3}$. Without a phase difference, a non-topological gap appears for low $g_0$ until eventually closing and remaining closed for larger $g_0$. With a phase difference, the non-topological gap remains but becomes a gapped topological state with $|\mathcal{C}|= 4$ as $g_0$ increases.

\begin{figure}[h]
    \centering
\includegraphics[width=.5\textwidth]{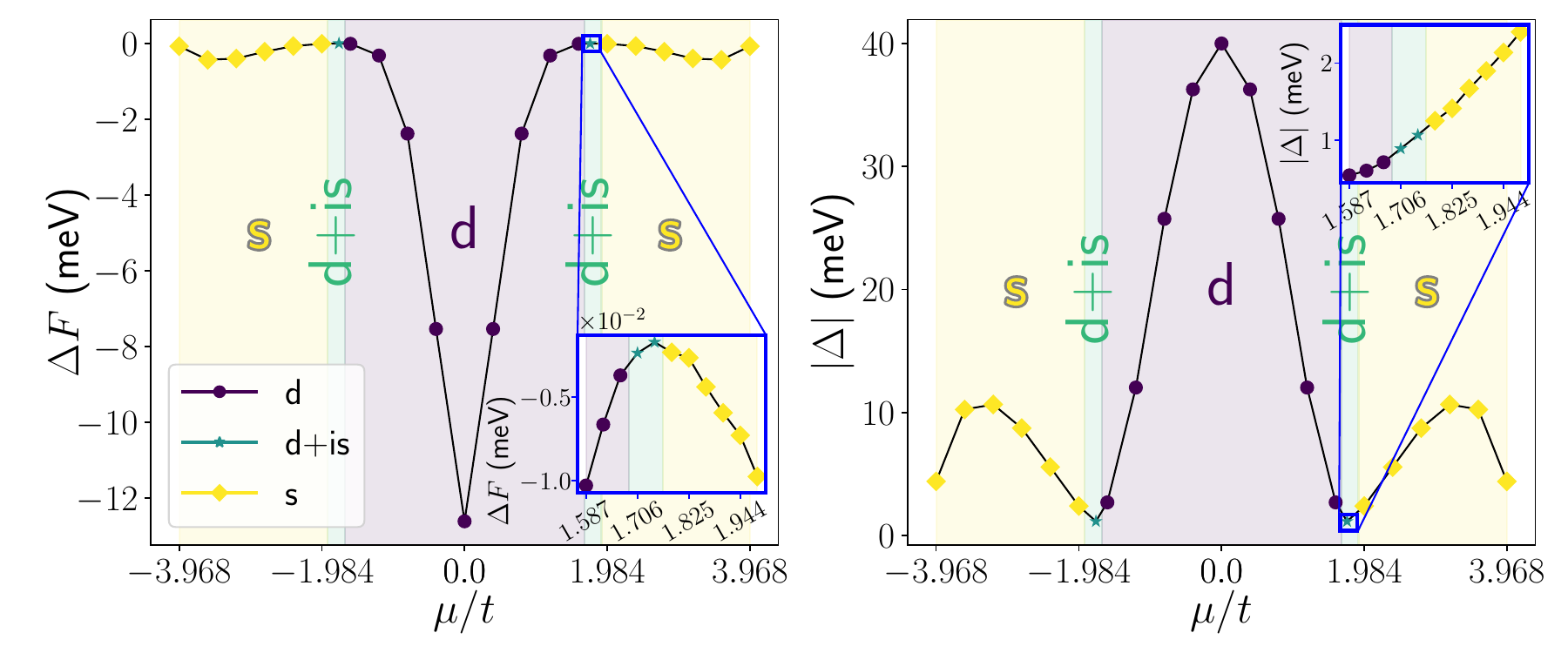}
    \caption{{\bf Single layer free energy and order parameter amplitude}. Shown above on the right is the change in free energy between the superconducting state and the normal state for varying chemical potential. The left shows the corresponding order parameter amplitude for varying chemical potential. Insets show the region where TRSB occurs leading to a $d+is$ order parameter. Due to particle-hole symmetry in the model, the diagram is symmetric across $\mu=0.0$ meV. }
    \label{fig:SI_single_layer_SC}
\end{figure}

\textit{Single-layer self-consistent solutions}---
To ascertain whether the moir\'{e} effects that accompany the $d$-wave order parameter are favorable, we self-consistently determined the order parameter sweeping over values of the chemical potential. We start with a single layer with the normal state Hamiltonian:
\begin{eqnarray}
H_{norm} = \sum_{\langle ij\rangle} \left( t c^\dagger_{i} c_{j}  + h.c.\right) - \mu\sum_{i}  n_i.
\label{eqn:single_layer_normal}
\end{eqnarray}
The results of varying the chemical potential are shown in Fig.~\ref{fig:SI_single_layer_SC}. Here, $\Delta F = F_{SC} - F_{N}$ where $F_{SC}$ is the free energy of the system in the superconducting state while $F_N$ is the free energy of the normal state. Until $|\mu/t| \lesssim 2$ the order parameter is $d$-wave until a transition occurs where the order parameter is a mixed $d+is$ state before becoming fully extended $s$-wave. One would therefore expect that for the twisted bilayers at small tunneling strength, that the order parameter for the layers will stem from a $d$-wave state about $\mu = 0.0$ meV.

\begin{figure}[h]
    \centering
\includegraphics[width=.5\textwidth]{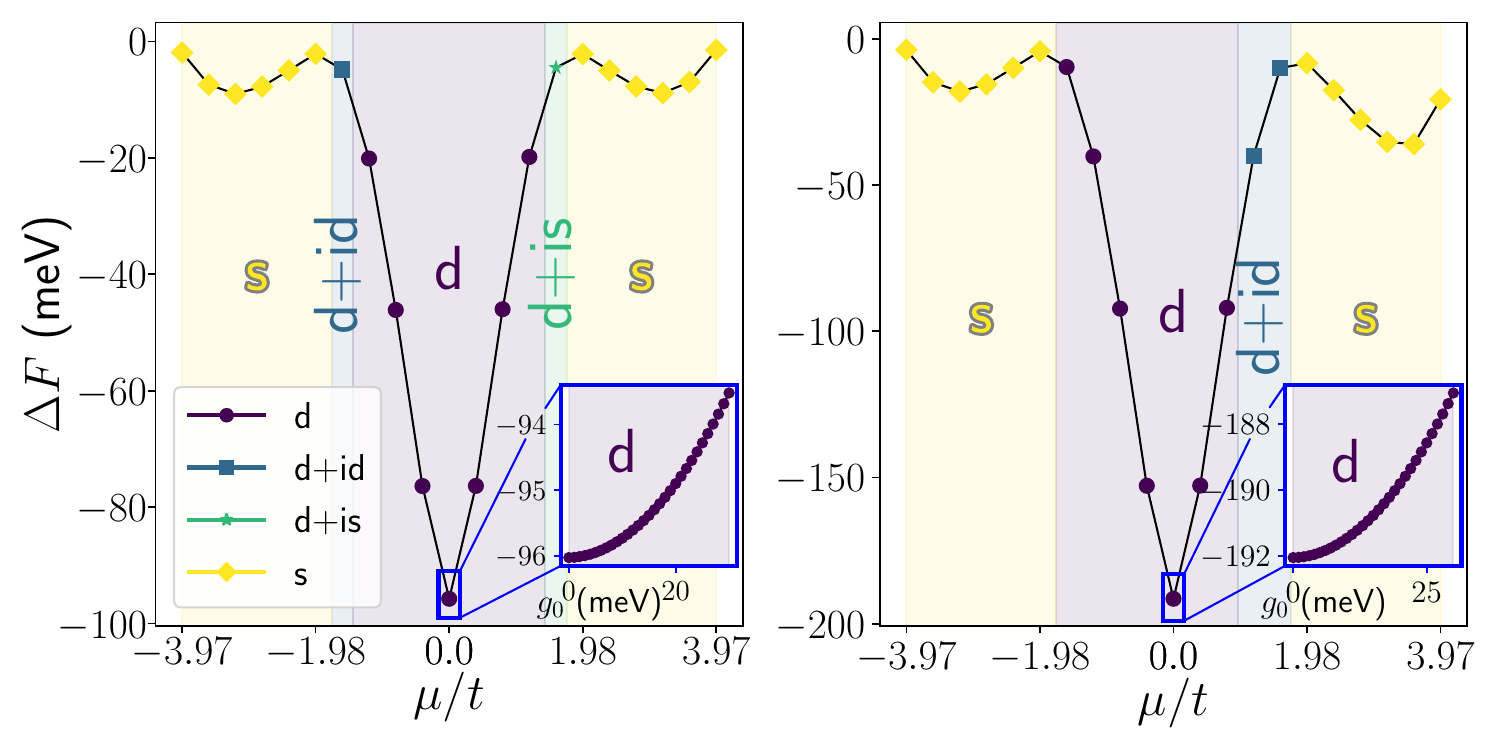}
    \caption{{\bf Self-consistent solution for twisted bilayers}. Shown are plots for $\Delta F$ vs $\mu$  for $g_0 = 10$ meV with the order parameter marked along the curves. Inset in each figure is a plot of $\Delta F$ vs $g_0$ for $\mu_c = 0$ meV boxed in blue on the main figure. Left are results for the $\theta_{1,2}$ twisted structure while right are results for $\theta_{1,3}$. For both, topological superconductivity can arise far from $\mu_c$, but varying $g_0$ at $\mu_c$ yields only a $d$-wave solution.}
    \label{fig:SC_with_twist}
\end{figure}

\textit{Bilayer self-consistent solutions}---
From the lattice model Eq.~\ref{eqn:Hlat}, we consider a self-consistently determined order parameter for small $g_0$ ($g_0 = 10$ meV) at varying chemical potential. Considering the results in Fig.~\ref{fig:SI_single_layer_SC}, we expect that the phase diagram remains consistent with the possibility of TRSB order parameters near the regions of $d+is$. Indeed, as shown in Fig.~\ref{fig:SC_with_twist} for $\theta_{1,2}$ and $\theta_{1,3}$, $d+id$ can form but far from $\mu_c$. In fact, near $\mu_c$, the $d$-wave state survives even as $g_0$ is varied up to $g_0= 40$ meV when fixed at $\mu_c$. For $\mu_{c,2}$ at $\theta_{1,3}$, it is likely within the region where $s$-wave is dominant and therefore is likely not feasible to demonstrate moir\'{e} effects.

\textit{Fitting BSSCO one band model}---
The dispersion of the one-band model can be expressed as:~\cite{Markiewicz2005}
\begin{equation}
\begin{gathered}
    E(\bm{k}) = -2t_{1} \big( \cos k_x a + \cos k_y a) - 4 t_{2} \cos k_x a \cos k_y a  \\
    - 2 t_3 \big( \cos 2k_x a + \cos 2 k_y a \big) 
    - 2 t_4 \big( \cos 2 k_x a \cos k_y a  \\ 
    + \cos 2 k_y a \cos k_x a \big)-\mu,
\end{gathered}
\end{equation}
where $a=5.4$ \AA~is the lattice constant, $\mu$ is the chemical potential, and $t_N$ are the $N$-nearest neighbor hopping strengths where $N=1,2,3,4$.  Taking dressed parameters to fit the model: $t_1 = 126$ meV, $t_2 = -36$ meV, $t_3 = 15$ meV, and $t_4 = 3.0$ meV~\cite{Markiewicz2005}. In the 2nd BZ, umklapp processes that would appear to be caused by $\theta_{c,2}$ are actually those of $\theta_{c}$ as the twist angle required is beyond $45^\circ$ (i.e. $48.9^\circ$). This causes the order parameter of one layer to acquire an overall phase shift of $\pi$. To ensure $\theta_{c,2}$ is less than $45^\circ$, we must go to the 3rd BZ where we find $\theta_{c,2}=30.5^\circ$. Therefore, $\theta_c $ is expected to occur for $40.1^\circ$ and $48.9^\circ$, while the umklapp process becomes negligible for $\theta_{c,2}$ as it occurs in the 3rd BZ.

\bibliography{main}

\clearpage
\onecolumngrid  

\setcounter{section}{0}
\renewcommand{\thesection}{\Roman{section}}
\renewcommand{\thefigure}{S\arabic{figure}}
\renewcommand{\theequation}{S\arabic{equation}}
\setcounter{secnumdepth}{3}
\setcounter{figure}{0}
\setcounter{equation}{0}

\setcounter{page}{1}  

\title{Supplemental Material for: ``Moir\'e-induced gapped phases in twisted nodal superconductors"}
\author{Kevin P.~Lucht}
\email{kpl55@rutgers.edu}
\affiliation{%
 Department of Physics and Astronomy, Center for Materials Theory,
Rutgers University, Piscataway, NJ 08854, USA
}%
\author{J.~H.~Pixley}
\affiliation{%
 Department of Physics and Astronomy, Center for Materials Theory,
Rutgers University, Piscataway, NJ 08854, USA
}%
\affiliation{
Center for Computational Quantum Physics, Flatiron Institute, 162 5th Avenue, New York, NY 10010
}%
\author{Pavel A.~Volkov}%
\affiliation{%
Department of Physics, University of Connecticut, Storrs, Connecticut 06269, USA
}%

\begin{center}
    \textbf{ \large Supplemental Material for: ``Moir\'e-induced gapped phases in twisted nodal superconductors"}\\[10pt]
    
    Kevin P.~Lucht\textsuperscript{1,*},
    J.~H.~Pixley\textsuperscript{1,3},
    Pavel A.~Volkov\textsuperscript{2}\\[4pt]
    
    \textsuperscript{1} \textit{Department of Physics and Astronomy, Center for Materials Theory,
Rutgers University, Piscataway, NJ 08854, USA}\\
    \textsuperscript{2} \textit{Department of Physics, University of Connecticut, Storrs, Connecticut 06269, USA}\\
    \textsuperscript{3} \textit{Center for Computational Quantum Physics, Flatiron Institute, 162 5th Avenue, New York, NY 10010}\\[10pt]
\end{center}

\section{Lattice Model of Twisted Nodal Superconductors}
\label{appendix:quant}
The Hamiltonian for the lattice model is given by
\begin{eqnarray}
H_{lat} = \sum_{\langle ij\rangle,l} \left( t c^\dagger_{i,l} c_{j,l}  + h.c.\right) - \sum_{i} \mu n_i   + \sum_{i,j < \Lambda} \left( g(r_{ij})c^\dagger_{i,1} c_{j,2}  + h.c. \right),
\label{eqn:num_real_space}
\end{eqnarray}
where $i,j$ are site indexes for each layer, $l$ is the layer index, $t$ is the nearest neighbor hopping strength, $\Lambda$ is a radial cut-off distance to limit the range of the tunneling. The cut-off distance is chosen to include the nearest 100 sites. The term $g(r_{ij})$ is the tunneling strength given as
\begin{equation}
    g(r_{ij}) = g_0 e^{-\frac{{\sqrt{r^2_{ij} - d^2} - d}}{\rho}},
    \label{eq:SI_tun_strength}
\end{equation}
where $g_0$ is the bare tunneling strength,  $r_{ij}$ is the distance between sites in the $xy$-plane, $d$ is the interlayer spacing, and $\rho$ is the radial extend of the interlayer tunneling.

Without self-consistency, we can include a term  $\sum_{\langle ij\rangle,l} \Delta_{ij} \left( c^\dagger_{i,l} c^\dagger_{j,l} + h.c. \right) $ where  $\Delta_{ij} $ is the hopping-dependent strength of the order parameter taken as
\begin{equation}
    \Delta_{ij} = \begin{cases}
        \phantom{-}\Delta_0/2 \quad \hat{x}, \\
        -\Delta_0/2 \quad \hat{y},
    \end{cases}
\end{equation}
to provide $d$-wave pairing. If an interlayer current is incorporated, the order parameter has layer-dependence $\Delta_{ij,l}$ where 
\begin{equation}
    \Delta_{ij,l} =  (-1)^l e^{i \varphi/2} \Delta_{ij},
\end{equation}
where $l=1,2$ signifies the layer index. We take $\varphi = \frac{\pi}{2}$ such that the phase difference between layers is $\Delta\varphi = \frac{\pi}{2}$.

To determine the superconducting order parameter self-consistently, we use an attractive nearest-neighbor interaction of the form
\begin{eqnarray}
    H_{int}  = -\sum_{\langle ij\rangle,\sigma\sigma'}  V n_{i\sigma}n_{j\sigma'},
\end{eqnarray}
where $V$ is the attractive Hubbard strength, $i,j$ are site indices on the square lattice, and $\sigma,\sigma'$ represent spin states $\sigma \in \{ \uparrow,\downarrow\}$. By applying a Hubbard-Stratonovich transformation, the interacting Hamiltonian can be expressed as
\begin{equation}
H_{I}(\bm{k}) = \sum_{\bm{k}} \left( \bar{ \Delta}_{\bm{k}} c_{-\bm{k}\downarrow}c_{\bm{k} \uparrow} + h.c. \right),
\end{equation}
where $\Delta_{\bm{k}} = \Delta_0 \gamma_{\bm{k}}$ where $\Delta_0$ is the amplitude of the order parameter and $\gamma_{\bm{k}}$ is a momentum dependent form-factor.
For the singlet channel, the form-factor is found to be either $d$-wave or extended $s$-wave.

To study twisted geometries, we focus on commensurate twist angles where periodic moir\'{e} superlattices are constructed. Commensurate twist angles correspond to a `twist' vector composed of two integer values $\bm{v} = (n,m)$ where $n,m \in \mathbb{Z}$ which corresponds to $2q = 2(n^2 + m^2)$ atoms in the moir\'{e} unit cell, or $q$ atoms per layer, and a twist angle of 
\begin{equation}
    \theta_{nm} = 2 \arctan \frac{n}{m}.
    \label{eq:twistnm}
\end{equation}
This angle corresponds to rotating the top and bottom layers by $\pm\theta_{nm}/2$, respectively. 

\section{BCS Decoupling of Nearest-Neighbor Interaction}
We follow a standard derivation for determining the form factors for the order parameter on a square lattice, see Ref.~\cite{Coleman_2015} for more details.
For a single square lattice, starting with a density-density interaction and assuming some attractive pairing potential, $V$, we have the real-space Hamiltonian:

\begin{equation}
    \hat{H}_I = -V \sum_{\langle i j \rangle} \hat{n}_{i\uparrow} \hat{n}_{j\downarrow}.
\end{equation}
Once Fourier transformed and assuming a BCS instability, we arrive at
\begin{eqnarray}
    \hat{H}_I(k)  = -V\sum_{\bm{k},\bm{k}',\delta_i} e^{i(\bm{k}'-\bm{k})\cdot \bm{\delta}_i} c^\dagger_{k\uparrow}c^\dagger_{-k\downarrow}c_{-k'\downarrow}c_{k'\uparrow},
\end{eqnarray}
where $\bm{\delta}_i$ correspond to NN vectors with the distance set to 1. The sum over NN gives
\begin{eqnarray}
    \hat{H}_I(k)  = -2V\sum_{k,k'} \left( \cos(k'_x-k_x) +\cos(k'_y-k_y) \right)   c^\dagger_{k\uparrow}c^\dagger_{-k\downarrow}c_{-k'\downarrow}c_{k'\uparrow}.
\end{eqnarray}
After inserting a trig identity, and ignoring the unphysical odd solutions for singlet pairing, we can decompose the even pairing potential into
\begin{eqnarray}
    \hat{H}_I(k)  = -V\sum_{k,k'} \left[ \left( \cos k_x +\cos k_y \right) \left( \cos k'_x +\cos k'_y \right)  + \left( \cos k_x - \cos k_y \right) \left( \cos k'_x - \cos k'_y \right) \right] c^\dagger_{k\uparrow}c^\dagger_{-k\downarrow}c_{-k'\downarrow}c_{k'\uparrow} \\
    = \sum_{k,k'} \left( V_s\gamma_s(k)\gamma_s(k') + V_d\gamma_d(k)\gamma_d(k')  \right) c^\dagger_{k\uparrow}c^\dagger_{-k\downarrow}c_{-k'\downarrow}c_{k'\uparrow},
\end{eqnarray}
where $V_s = V_d = -V$, $\gamma_s(k) = \cos k_x + \cos k_y$, and $\gamma_d(k) = \cos k_x - \cos k_y$.
Performing a Hubbard-Stratonovich transformation on the interaction of the form $H_I = -\frac{1}{V}A^\dagger A$ where
\begin{equation}
    \hat{A}= \sum_{k} \gamma_k c_{-k\downarrow}c_{k\uparrow}, \quad     \hat{A}^\dagger= \sum_{k} \gamma_k c^\dagger_{k\uparrow}c^\dagger_{-k\downarrow},
\end{equation}
we find:
\begin{equation}
\hat{H}_I = \sum_i \left[ \sum_{k} \gamma_{i,k}(\bar{\Delta}_i(k) c_{-k\downarrow}c_{k\uparrow} + h.c.) + V \bar{\Delta}_i(k){\Delta}_i(k)  \right],
\end{equation}
where $i = s,d$ and $\Delta_i(k) = \Delta_i \gamma_i(k)$. Here, $\Delta_i$ is allowed to be imaginary in order to account for time-reversal symmetry broken order parameters.

\section{Self-Consistent Methods}

To get the free energy, we first express the effective action as~\cite{Can2021} 

\begin{equation}
    S_{eff} = - \sum_{\bm{k},n} \mathrm{Tr}\log[\mathcal{G}(\bm{k},i\omega_n)^{-1}] + \frac{\beta \mathcal{N}}{V}\sum_{i,j \in U.C.} \Delta_{ij,l}\bar{\Delta}_{ij,l},
\end{equation}
where the Matsubara Green’s function is $\mathcal{G}(\bm{k},i\omega_n)^{-1} = -(-i\omega + h_{\bm{k}})^{-1}$, $\mathcal{N}$ is the number of layers, $V$ is the attractive interaction strength. Using standard contour integration techniques to evaluate the Matsubara summation, we arrive at:
\begin{equation}
    S_{eff}  = -\mathrm{Tr}[\ln(1+e^{-\beta h_{\bf{k}}}) ] + \frac{ \beta\mathcal{N}}{V}\sum_{i,j \in \Lambda} \Delta_{ij,l}\bar{\Delta}_{ij,l}.
\end{equation}
Making use of the identity $U_{\bm{k}}U^\dagger_{\bm{k}}= \mathbb{I}$ where $U_{\bm{k}}$ is the unitary matrix that diagonalizes the Bloch Hamiltonian as $U^\dagger_{\bm{k}} h_{\bm{k}} U_{\bm{k}} = E_{\bm{k}}$, we have 
\begin{equation}
    S_{eff}  = -\mathrm{Tr}[\ln(1+U_{\bm{k}}e^{-\beta E_{\bf{k}}}U^\dagger_{\bm{k}}) ] + \frac{ \beta\mathcal{N}}{V}\sum_{i,j \in \Lambda} \Delta_{ij,l}\bar{\Delta}_{ij,l}.
\end{equation}
The corresponding Free energy, $\mathcal{F}_{eff}$, can be written as
\begin{equation}
    \mathcal{F} = -\frac{ln \mathcal{Z}}{\beta} =  -\frac{1}{\beta}\mathrm{Tr}[\ln(1+U_{\bm{k}}e^{-\beta E_{\bf{k}}}U^\dagger_{\bm{k}}) ] + \frac{\mathcal{N}}{V}\sum_{i,j \in \Lambda} \Delta_{ij,l}\bar{\Delta}_{ij,l}.
\end{equation}
To determine the Free energy of the ground state, we take the limit $\beta \rightarrow \infty$. Making use of the fact that
\begin{equation}
    \ln(1+U_{\bm{k}}e^{-\beta E_{\bf{k}}}U^\dagger_{\bm{k}})  = \ln(U_{\bm{k}}(1+e^{-\beta E_{\bf{k}}})U^\dagger_{\bm{k}})  \approx  - \frac{1}{2} \beta E_{\bm{k}} + \frac{1}{2}\beta | E_{\bm{k}}|,
\end{equation}
in this limit. Taking into account the particle-hole symmetry of the energy eigenvalues, the free energy simplifies to 
\begin{equation}
    \mathcal{F}(\beta \rightarrow \infty) \approx -\frac{1}{2}\sum_{n\bf{k}} |E_{n\bf{k}}|+ \frac{ \mathcal{N}}{V}\sum_{i,j \in \Lambda} \Delta_{ij,l}\bar{\Delta}_{ij,l}.
\end{equation}
where the factor of $1/2$ avoids double-counting of energy eigenstates and $n$ is the band index.

To self-consistently determine the order parameters from the free energy, we employ a Broyden method~\cite{Broyden1965} to approximate the Jacobian of the variables, in this case $\Delta_{ij}$ and $\bar{\Delta}_{ij}$. Our goal is to determine the solution to the system of equations, $\bm{f}(\Delta,\bar{\Delta})= \bm{0}$ where each expression is given by
\begin{equation}
    \bm{f}(\Delta_{ij},\bar{\Delta}_{ij}) = \frac{1}{2} \Big( \frac{\partial \mathcal{F}(\beta \rightarrow \infty)}{\partial \Delta_{ij}}  + \frac{\partial \mathcal{F}(\beta \rightarrow \infty)}{\partial \bar{\Delta}_{ij}} \Big).
\end{equation}
To minimize these set of functions, we provide an initial ansatz to $\Delta_{ij}$ and take the initial Jacobian as $J_n = \mathbb{I}$ where $n$ denotes the iterative step. To scan over all possible ground state solutions of the order parameter, we consider all possible forms of $s$,$d+is$,$s+is$, $d+id$, or $d$. We then perform the recursive steps:
\begin{enumerate}
     \item $\delta_n \Delta_{ij} = - J_n \bm{f}(\Delta_{ij,n},\bar{\Delta}_{ij,n})$,
     \item $\Delta_{ij,n+1} = \alpha\Delta_{ij,n} +  (1-\alpha)\delta_n\Delta_{ij} $,
     \item $\delta \bm{f} = \bm{f}(\Delta_{ij,n+1},\bar{\Delta}_{ij,n+1}) - \bm{f}(\Delta_{ij,n},\bar{\Delta}_{ij,n})   $, \\
     \item $\bm{J}_{n+1} = \bm{J}_{n}  + \frac{ (\delta_n \Delta_{ij}  - \bm{J}_n \delta \bm{f} ) \delta_n \Delta_{ij}^T }{\delta_n \Delta_{ij}^T \delta_n \Delta_{ij}}  $,
\end{enumerate}
where $\alpha$ is a mixing parameter to mix new and old solutions to $\Delta_{ij,n+1}$. At fixed step intervals, $J_n$ is reset to $\mathbb{I}$ to restart the recursion process to improve convergence. We take a convergence criteria for all parameters of $\delta \bm{f} = 10^{-12}$. To ensure the free energy is minimized, we also compare the free energy for each recursive step. If the free energy increases, we backtrack $\Delta_{ij,n+1}$ by including a recursive sub-routine by taking $\Delta_{ij,n+1} =\Delta_{ij,n+1} - \gamma( \Delta_{ij,n+1} - \Delta_{ij,n})$ where $\gamma$ is a parameter which scales $\Delta_{ij,n+1} $ back towards the previous step. $\gamma$ is scaled close to $0$ towards $1$, comparing the free energy along the way. The lowest free energy is taken and the original recursion continues. Once a solution converges, the magnitude (if the anstaz is taken as $s$ or $d$) or phase (if the ansatz is $s+is$, $d+is$, or $d+id$) is slightly shifted in small steps and the free energy is compared. If the free energy is smaller, the recursion is restarted with the new shifted parameters taken as the starting ansatz. After compiling all possible order parameter ansatzes, the minimum free energy solution is selected as the ground state solution.

\section{Determining Topological Phases}
To determine the topology in our lattice model, we first compute the Berry curvature, $\Omega(\bm{k})$, over the full Brillouin zone of the commensurate moir\'{e} superlattice where
\begin{equation}
\Omega(\bm{k}) = 2 \textrm{Im}\sum_{m,n} \frac{\langle n | \hat{v}_x |m \rangle \langle m| \hat{v}_y | n \rangle}{(E_m -E_n)^2},
\end{equation}
where $\hat{v}_i = \frac{\partial \hat{H}_{MFT}(\bm{k})}{\partial k_i}$, where $\hat{H}_{MFT}(\bm{k})$ is the Fourier transformed Hamiltonian from Eq.~\ref{eqn:Hlat} and $m,n$ index the eigenstates $|m\rangle$, $| n \rangle$ of the Hamiltonian with corresponding energy eigenvalues $E_m$ and $E_n$, respectively. The Chern number is then computed by summing $\Omega(\bm{k})$ over the momentum grid,
\begin{equation}
    \mathcal{C} = \frac{1}{2\pi} \frac{1}{N_k}\sum_{\bm{k}} \Omega(\bm{k}),
\end{equation}
where $N_k$ is the number of points in the k-mesh which acts as the measure of the numerical integral. When a $d+id$ order parameter is introduced, the resulting Chern number is the sum of the nodes in each valley, $|\mathcal{C}| = 4$. However, if intervalley tunneling is dominate, the level repulsion results in a gapped non-topological state with $|\mathcal{C}| = 0$. Once intravelly tunneling dominates once again, a the gapped state is a returns to the expected topological $|\mathcal{C}| = 4$ state.


\section{Continuum Model of Twisted Nodal Superconductors}
\label{appendix:quantA}

For a single-layer nodal superconductor, the low energy continuum model is governed by the BdG Dirac cone formed about a node at momentum $\bm{K}_N$ from the origin:
\begin{equation}
    \mathcal{H}_1 = \sum_{\bm{k}} \Phi_{1,\bm{k}}^\dagger \left( v_F k_\parallel \tau_3 + v_\Delta k_\perp \tau_1  \right) \Phi_{1,\bm{k}},
    \label{eqn:single_layer_Ham}
\end{equation}
where $\Phi^\dagger_{1,\bm{k}}$ is a Balian-Werthammer (B-W) spinor, $v_F$ ($v_\Delta$) is the velocity of the normal state (order parameter), and $k_\parallel$ ($k_\perp$) is the momentum local to a node and is oriented parallel (perpendicular) to $\bm{K}_N$. The Pauli matrices $\tau_i$ where $i=1,2,3$ are matrix representations of Gor'kov-Nambu space and $\tau_0$ is treated as the identity matrix. When a second layer is stacked on top of layer 1 to form a homobilayer, the Hamiltonian for a valley, $\mathcal{H}_2$, when the lattices are aligned is:
\begin{equation}
    \mathcal{H}_2 = \sum_{\bm{k}} \Phi_{\bm{k}}^\dagger \left( v_F k_\parallel \tau_3 + v_\Delta k_\perp \tau_1  + t \sigma_1\tau_3 \right) \Phi_{\bm{k}},
    \label{eqn:two_layer_Ham}
\end{equation}
where $\Phi^\dagger_{\bm{k}} = [ \Phi^\dagger_{1,\bm{k} },\Phi^\dagger_{2,\bm{k} }] $ is a B-W spinor composed of components from each layer's B-W spinor, $\Phi^\dagger_{l,\bm{k}}$, where $l =1,2$ labels the layer index, and $t = t_{K_N}/\Omega$ is the normalized tunneling strength given by the tunneling strength at a node, $t_{K_N}$, divided by a volume form factor, $\Omega$ of the Brillouin zone (BZ). This Hamiltonian is essentially two copies of the single layer Hamiltonian, Eq.~\ref{eqn:single_layer_Ham}, with the only new addition coming from the hopping between layers.

Once a small twist angle of $\theta/2$ and $-\theta/2$ is applied to layer 2 and 1, respectively, the low energy continuum model in a given valley becomes:~\cite{pavel_article_2023,pavel_letter_2023}
\begin{equation}
    \mathcal{H} = \sum_{\bm{k}} \Phi_{\bm{k}_{\theta}}^\dagger \left( v_F k_\parallel \tau_3 + v_\Delta k_\perp \tau_1  + t\sigma_1 -\alpha t \tau_1 \sigma_3 \right) \Phi_{\bm{k}_\theta},
    \label{eqn:twisted_bilayer_Ham}
\end{equation}
where $\Phi^\dagger_{\bm{k}_\theta} = [ \Phi^\dagger_{1,\bm{k}-\bm{Q}_N/2 },\Phi^\dagger_{2,\bm{k}+\bm{Q}_N/2 }] $ is a B-W spinor composed of the rotated layer components from each layer and $Q_N= (\hat{z} \times \bm{K}_N) \cdot \theta$ is the magnitude of the momentum of shifting the Dirac node along $\hat{k}_\perp$ due to the twisting by angle $\theta$. The new parameter $\alpha = \delta_0/t$ is the normalized angle where $\delta_0 = v_\Delta Q_N/2$.

Furthermore, by introducing an interlayer current, a phase difference between the order parameters of the two layers appears due to the Josephson-current relationship:
\begin{equation}
    I(\Delta \varphi) = I_c \sin \Delta \varphi,
\end{equation}
where $\Delta\varphi$ is the phase difference between the order parameters in the two layers and $I_c$ is the magnitude of the Josephson current. This phase difference essentially transforms the pairing component in Nambu space, i.e. $\tau_1 \rightarrow \cos \varphi/2 \tau_1 - \sin\varphi/2 \tau_2 \sigma_3$, where the $\sigma_3$ reflects the phase is $\pm \varphi/2$ for the top and bottom layer, respectively, such that the phase difference is $\varphi$. Performing this transformation can be thought of as a rotation in Nambu space permitted by broken mirror symmetry by twisting, which introduces a term to the Hamiltonian that breaks time-reversal symmetry (i.e. $\alpha t \tau_2)$. Consequently, a topological gap opens in the BdG spectrum, forming a topological superconductor with Chern number $|\mathcal{C}|=N_v$ where $N_v$ is the number of valleys~\cite{pavel_article_2023,pavel_letter_2023}.


\section{Continuum Model at Large Twist Angle} 

\label{AppA}
For two layers which are at a large twist angle, we can treat this as two nodes in a valley separated by some large momenta along the perpendicular by $\pm\delta_0$ where $\delta_0  >> t$. In this way, we can treat $t$ as a perturbation to the modified expression of Eq.~\ref{eqn:two_layer_Ham} written as:

\begin{equation}
    \mathcal{H}_{\bm{k}} = \sum_{\bm{k}} \Psi^\dagger_{\bm{k}} \left(  \xi \tau_3 + \delta \tau_1 + \delta_0 \tau_1 \sigma_3 \right)  \Psi_{\bm{k}}.
\end{equation}
This expression can also be extended in the case of an interlayer current through the system by including a phase of $e^{\pm i \varphi/2}$ to the top and bottom layer, respectively. We can now proceed to include a tunneling term, represented by $t\tau_3\sigma_1$, which we will treat as a perturbation about zero energy. Zero energy is achieved at the position of each node, i.e. $\xi = 0$ and $\delta = \pm \delta_0$ which translate to zero energy eigenstates spanned by
\begin{equation}
    | \psi_{1,+} \rangle = (1,0,0,0)^T, \quad | \psi_{2,+} \rangle = (0,1,0,0)^T,
\end{equation}
for the nodes positioned at $\xi = 0$ and $\delta = \delta_0$, and likewise
\begin{equation}
    | \psi_{1,-} \rangle = (0,0,0,1)^T, \quad | \psi_{2,-} \rangle = (0,0,1,0)^T,  
\end{equation}
for nodes positioned at $\xi = 0$ and $\delta = -\delta_0$. 

We will now focus on an individual node introduce the tunneling strength between nodes. When positioned at one node, the energy scale to tunnel to the other node is therefore $2\delta_0$ and leads to high energy eigenstates being a linear combination of the zero energy eigenstates of the other node. Treated at the level of second order perturbation theory, we acquire the contribution expressed in first quantized form as
\begin{equation}
\delta H_{\pm} =
 \pm \frac{t^2}{2\delta_0} \tau_1,
\end{equation}
for layers shifted by $\pm \delta_0$, respectively.
The local physics for the full valley with nodes shifted by $\pm \delta_0$ can then be represented by the Hamiltonian in first quantized form as
\begin{equation}
H_{\pm,eff}(\bm{k}) = \begin{pmatrix}
    \xi & \delta \pm \frac{t^2}{2\delta_0} \\
    \delta \pm  \frac{t^2}{2\delta_0}  & -\xi  
\end{pmatrix}.
\label{eqn:Heff_tun}
\end{equation}
In the case of an interlayer current, a gauge transformation can be applied to the creation and annihilation operators which repositions the phase onto the shift $\delta_0$ in the high-energy sector. Consequently, the eigenstates of the high-energy sector also acquires the phase so that when reapplying second order perturbation theory, we find the effective Hamiltonian for the full valley
\begin{equation}
H_{eff,\pm}(\bm{k},\varphi) = \begin{pmatrix}
    \xi & \delta \pm \frac{t^2}{2\delta_0}e^{\mp i\varphi/2} \\
    \delta \pm \frac{t^2}{2\delta_0}e^{\pm i\varphi/2}  & -\xi  
\end{pmatrix}.
\label{eqn:Heff_tun_phase}
\end{equation}

\section{Chern Number of Effective Hamiltonian for $t << \delta_0$}
\label{AppB}

Corresponding to each block, we can determine the Chern number for a two band model. In this case, we can express each block  in Eq.~\ref{eqn:Heff_tun_phase} as
\begin{equation}
    H_{eff,\pm}(\bm{k},\varphi) = \xi \tau_3 + \left( \delta \pm \frac{t^2}{2\delta_0}\cos(\varphi/2) \right) \tau_1 -  m \sin(\varphi/2) \tau_2,
\end{equation}
where the topological mass $m = \frac{t^2}{2\delta_0}  $. Expressed as a vector, we can write
\begin{equation}
    H_{eff,\pm}(\bm{k},\varphi) = \bm{d}_\pm(\bm{k},\varphi) \cdot \bm{\tau},
\end{equation}
where the components of $\bm{d}_\pm(\bm{k},\varphi)$ for each respective Hamiltonian are therefore
\begin{eqnarray}
    d_{3,\pm} = v_F k_\parallel, \quad
    d_{2,\pm} = -\frac{t^2}{2\delta_0}\sin \varphi/2, \quad 
    d_{1,\pm} = \left( \delta \pm \frac{t^2}{2\delta_0}\cos\varphi/2 \right).
\end{eqnarray}
Each Hamiltonian will ultimately provide a Chern number, with a gap corresponding to $\Delta =|\frac{t^2}{\delta_0}\sin\varphi|$. A few remarks regarding the resulting topology, for one the mass within a valley is the same and will yield to $|\mathcal{C}| =1$ inside a valley. Secondly, the gap corresponds to the tunneling strength which when compared to the other valley from the 1-st shell will be smaller in magnitude (and opposite in sign) which can modify the net Chern number (and hence the topology). Following the standard procedure outlined in Ref.~\cite{bernevig2013topological}, one can show the Berry curvature is
\begin{equation}
    \Omega_{xy,\pm}(\varphi)
    = v_\Delta v_F\frac{t^2}{2\delta_0}\frac{\sin \varphi}{2E^3},
\end{equation}
where $E$ is the energy of the occupied BdG band and $x,y$ represent the directions parallel and perpendicular to the unrotated position of a BdG node, respectively. Integrating to compute the Chern number, we find
\begin{eqnarray}
\mathcal{C}_\pm  = \frac{1}{2\pi} \int d \bm{k}\Omega_{xy} = \frac{1}{4}\frac{t^2}{2\delta_0}\sin\varphi\int_0^\infty\frac{ dx }{(x+\left(\frac{t^2}{2\delta_0}\right)^2 \sin^2\varphi)^{3/2}}
 = \frac{1}{2} \textrm{sgn}( \sin\varphi),
\end{eqnarray}
where we used polar coordinates and a substitution of variables to get the integrand.

\section{Tunneling Strength}

To understand the formation of the non-topological gap, we will first consider how $t_{K_N^{(2)}}$ exceeds $t_{K_N^{(1)}}$ using the parameters of our lattice model to fit a Bistritzer-MacDonald continuum model of the tunneling strength in Eq.~\ref{eqn:tunneling_form_factor}. In particular, we are looking for $t_{|\bm{K}_N^{(1)}|}$ and $t_{|\bm{K}_N^{(2)}|} \equiv t_{|\bm{K}_N^{(1)}+\bm{G}|} $. We can achieve this by Fourier transforming the real-space tunneling strength given in Eq.~\ref{eq:num_tun_strength},

\begin{figure}[h]
    \centering
\includegraphics[width=.45\textwidth]{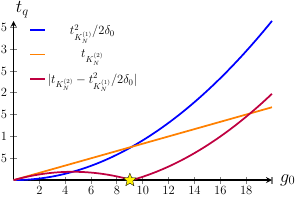}
    \caption{ Tunneling strength. Shown are the the tunneling strength $t_q$ for the intravalley process, $t_{K_N^{(1)}}^2/\delta_0$, and 1-st umklapp scattering process, $t_{K_N^{(2)}}$, versus bare tunneling strength $g_0$ in units of meV. Values were taken from the $\theta_{1,2}$ twisted structure. The gold star marks the transition from $t_{K_N^{(2)}}$ to $t_{K_N^{(1)}}^2/\delta_0$ at higher $g_0$. 
    }
    \label{fig:SM_tun_str}
\end{figure}

\begin{equation}
    t_{K_N^{(i)}} ={2\pi} \int dr r J_0(K_N^{(i)} r) g(r),
    \label{eq:SI_FT_tun}
\end{equation}
where $r$ is the radial distance in the plane, $J_0(z)$ is the Bessel function of first kind at order $0$, and $g(r)$ is the real space tunneling strength for $r$. Using the position of the BdG Dirac nodes and the reciprocal lattice vectors at a given twist angle, one can determine $t_{K_N^{(i)}}$ explicitly. For $t_{K_N^{(2)}}$ at $\mu_c$, this tunneling strength is set at the momentum that the nodes overlap in a higher BZ. The magnitude of $t_{K_N^{(1)}}$ is rescaled at large twist angles to $t_{K_N^{(1)}} \rightarrow t_{K_N^{(1)}}^2/\delta_0$. From the numerical parameters, $\delta_0 = |\bm{v}\cdot (\bm{K}_N^{\theta_{nm}/2} - \bm{K}_N^{-\theta_{nm}/2}) |$ where $\bm{v} = (v_\Delta,v_F)$ and $\bm{K}_N^{\theta_{nm}}$ is the position of the BdG Dirac node in the BZ rotated by an angle $\theta_{nm}$ (see Eq.~\ref{eq:twistnm}). For example, the results in Fig~\ref{fig:SM_tun_str} were found using $v_\Delta = 40$ meV, $|K_N^{(1)}| = 2.221 $ for intravalley tunneling and $|K_N^{(2)}| = 4.967 $ for umklapp tunneling. In our case, $\delta_0 = 40\times 2\times 0.702481$ meV as the intravalley nodes are separated strictly along $\hat{x}$. Here, we set the units of $[a]=1$ where $a = 5.4$ \AA~is the lattice constant of the monolayer. Setting $d = 12/5$ \AA~in Eq.~\ref{eq:SI_tun_strength}, we can numerically compute $t_{K_N^{(i)}}$. Note in our use of Eq.~\ref{eq:SI_FT_tun} our measure is scaled as $r\rightarrow r/a$. Computing the two quantities, we find $t_{K_N^{(2)}} = 0.0836$ meV and $t_{K_N^{(0}}^2/\delta_0  = 0.00913$ meV if we take $g_0 = 1$ meV. 

Due to the magnitude of $\delta_0$, $t_{K_N^{(2)}} > t_{K_N^{(1)}}^2/\delta_0$ for low $g_0$. Eventually, a critical value of $g_0$ is reached marked by the star in Fig.~\ref{fig:SM_tun_str} where for larger $g_0$, $t_{K_N^{(2)}}< t_{K_N^{(1)}}^2/\delta_0$. This exchange of dominant tunneling corresponds with the formation of the non-topological phase, and the transition from $|\mathcal{C}|=0$ to $|\mathcal{C}|=4$ with $\varphi\neq0$ shown in Fig.~\ref{fig:figure2_n1m3} and Fig.~\ref{fig:figure3_n1m3_muc2}. When performing the continuum modeling, Fig.~\ref{fig:cont_model} (c) uses the tunneling values determined for $\theta_{1,2}$ at $\mu_c$ when computing the effective mass.

\section{Continuum Model of Interlayer Tunneling}

We now consider coupling between valleys provided by the tunneling in higher Brillouin zones. The eigenstates in another valley we'll project to are $(0,0,0,1)$ and $(0,0,1,0)$, so the full result takes the general form:
\begin{equation}
\begin{gathered}
H_{2val} = 
\frac{1+\eta_3}{2} 
\left(\xi_1 \tau_3 +\left[\delta_1 + m_1\right] \cos (\varphi/2) \tau_1 + (m_1-\delta_1) \sin(\varphi/2) \tau_2\right)
+
\\
+
\frac{1-\eta_3}{2} 
\left(\xi_2 \tau_3 
+
\left[\delta_2 + m_2\right] \cos (\varphi/2) \tau_1 - (m_2-\delta_2) \sin(\varphi/2) \tau_2
\right)
+
t_{K_N^{(2)}}  \eta_1 \tau_3,
\end{gathered}
\end{equation}
where $\eta$ are Pauli matrices in valley space. In general, $|m_1|=|m_2| = (t_{K_N^{(1)}} )^2/(2 \delta_0)$, but their signs (as well as relations between $\xi_1,\delta_1$ and $\xi_2,\delta_2$) depend on the Fermi/gap velocities at the respective nodes. E.g., for the same valley, $m_2=-m_1$, while for next valley along the same layer's Fermi surface the sign of $m$ alternates

\subsection{Cases}

$\bullet$ Case 1 (adjacent valleys separating with twist): $\xi_2 = -\xi_1,\delta_2= \delta_1$,  $m_2=m_1$:

\begin{equation}
\begin{gathered}
H_{2val} = 
\xi \tau_3 \eta_3 +\delta \cos (\varphi/2) \tau_1 + m\cos (\varphi/2) \tau_1+ m \sin(\varphi/2) \tau_2 \eta_3
\\
-\delta \sin(\varphi/2) \tau_2 \eta_3
+
t_{K_N^{(2)}}  \tau_3 \eta_1
\end{gathered}
\end{equation}

The eigenvalues are:

\begin{equation}
\begin{gathered}
        E^2 = m^2 + (t_{K_N^{(2)}})^2 + \xi^2 + \delta^2 + 2 m \delta \cos\varphi \pm 
 2 t_{K_N^{(2)}}  |(m - \delta) \sin(\varphi/2)|
 =
 \\
 = \xi^2 + 
 (m+\delta)^2\cos^2\frac{\varphi}{2} + \left(|(m-\delta)\sin(\varphi/2)|\pm t_{K_N^{(2)}} \right)^2.
\end{gathered}
\end{equation}

For $\varphi=0$ there's always a $2t_{K_N^{(2)}} $ gap in the spectrum, which is attributed to avoided crossing of the nodal line by the hybridized Fermi surface. For finite $\varphi$, a topological transition where gap closes and reopens around one Dirac node occurs at  $2m\sin(\varphi/2) = t_{K_N^{(2)}} $. This effect is precisely captured in the numerical results in Fig.~\ref{fig:figure2_n1m3} (b).

$\bullet$ Case 2: (opposite valleys, i.e. $I$ and $[I \pm 2]$ mod $4$ where $I=1,2,3,4$): $\xi_2 = -\xi_1,\delta_2= - \delta_1$,  $m_2=-m_1$:

\begin{equation}
\begin{gathered}
H_{2val} = 
\xi \tau_3 \eta_3 +(\delta+m) \cos (\varphi/2) \tau_1 \eta_3+ (m-\delta) \sin(\varphi/2) \tau_2 
\\
+
t_{K_N^{(2)}}  \tau_3 \eta_1
\end{gathered}
\end{equation}

The eigenvalues are:

\begin{equation}
\begin{gathered}
        E^2 = m^2 + (t_{K_N^{(2)}}) ^2 + \xi^2 + \delta^2 + 2 m \delta \cos\varphi \pm 
 2 t_{K_N^{(2)}}  |(m + \delta) \cos(\varphi/2)|
 =
 \\
 =
 (m-\delta)^2\sin^2\frac{\varphi}{2} + \left(|(m+\delta)\cos(\varphi/2)|\pm t_{K_N^{(2)}} \right)^2.
\end{gathered}
\end{equation}

For $\varphi=0$ there is no gap for any $m,t_{K_N^{(2)}} $. The Dirac points are at $\delta = \pm t_{K_N^{(2)}} -m$.

For nonzero $\varphi$, on the contrary, there is always a gap, but it close and reopens at $2 m \cos(\varphi/2) = t_{K_N^{(2)}} $. Thus, a non-topological gap at low $m$ only opens when TRS is broken.

\section{Estimating $\theta_c$ and $t_{K_N^{(i)}}$}
Here we also estimate the critical twist angle by fitting to overdoped ($p > 0.16$) BSCCO. To determine the chemical potential, we rely on fitting to APRES measurements at various doping~\cite{Sobota2021} and match to the corresponding $T_c$~\cite{He2018}. To approximate $v_\Delta$, we take provided data of the antinodal gap, $\Delta_{AN}$, which is nearly identical to $v_\Delta$~\cite{He2018}. Taking a $d$-wave form factor, the dispersion of the order parameter is taken as:
\begin{equation}
    \Delta(\bm{k}) = \frac{v_\Delta}{2} \Big( \cos( k_x a) - \cos(k_y a) \Big),
\end{equation}
where $a$ is the lattice constant of $5.4$ \AA~which we normalize in our computations to $1$. To model the normal state, we adopt a phenomenological tight-binding model~\cite{He2018}:
\begin{equation}
    \epsilon_k = -2.52 \Big(\cos(k_x a) + \cos( k_y a) \Big) +2.12 \cos(k_x a) \cos( k_y a),
\end{equation}
which are renormalized with a momentum-dependent form factor:
\begin{equation}
    \lambda_{ee} = 1 + \sin^6\Big( \frac{\sqrt{k^2}}{8}a \Big).
\end{equation}
With these expressions and chemical potential values, we can determine the nodal positions and therefore the critical angles, $\theta_c$, which produce optimal umklapp tunneling strength. A table of used data and the resulting $\theta_c$ is provided in Tab.~\ref{tab:ARPES_fit}. Given $v_\Delta$ and $g_0$, we can now approximate the tunneling strengths. Here, we approximate $\delta_0 = v_\Delta | K^{\theta_c/2}- K^{-\theta_c/2}|$ for each filling $p$ and compute both $ t_{K_N^{(1)}}^2/\delta_0 $ and  $t_{K_N^{(2)}}$  for $g_0 = 1.0 $ meV which are also provided in Tab.~\ref{tab:ARPES_fit}.  Given the modeling for the normal state dispersion, the value of $t_{K_N^{(1)}}$ is amplified compared to the value provided in the main text, however, its suppression near optimal filling and $\theta_c$ remains consistent.

    \begin{table}
        \centering
        \begin{tabular}{c| cccccc}
            $p$ &$\mu$ (meV) & $T_c$ (K) & $v_\Delta$ (meV) & $\theta_c$ ( deg) & $ t_{K_N^{(1)}}^2/\delta_0 $ (meV) & $t_{K_N^{(2)}}$ (meV) \\ \hline
             0.16 & 17 & 95  & 43.9 & 37.0 & 0.0050 & 0.0015 \\
             0.18 & 27 & 92  & 42.4 & 37.1 & 0.0050 & 0.0015 \\
             0.20 & 34 & 80  & 29.1 & 37.2 & 0.0072 & 0.0015 \\
             0.21 & 41 & 74  & 26.1 & 37.2 & 0.0079 & 0.0015 \\
             0.22 & 47 & 64  & 13.5 & 37.3 & 0.015  & 0.0015 \\
             0.24 & 57 & 47  & 11.3 & 37.4 & 0.018  & 0.0015
        \end{tabular}
        \caption{Fitting data. Shown are for given filling fraction, $p$, the corresponding chemical potential, $\mu$, and the parameters used to determine $\theta_c$ and tunneling strengths.}
        \label{tab:ARPES_fit}
    \end{table}

\end{document}